\documentclass[aps,twocolumn,prb,showpacs,amsmath,superscriptaddress]{revtex4}
\usepackage{epsfig}
\usepackage{bm}
\def\be{\begin{equation}}
\def\ee{\end{equation}}
\def\bea{\begin{eqnarray}}
\def\eea{\end{eqnarray}}
\def\nn{\nonumber}
\def\upa{\uparrow}
\def\dwa{\downarrow}

\begin{document}
\title{Renormalization of impurity scattering in one-dimensional
  interacting electron systems in magnetic field}

\author{T. Hikihara} 

\affiliation{Division of Physics, Graduate School of Science, Hokkaido
  University, Sapporo, 060-0810, Japan}

\affiliation{Condensed Matter Theory Laboratory, RIKEN, Wako, Saitama
  351-0198, Japan} 

\author{A. Furusaki}

\affiliation{Condensed Matter Theory Laboratory, RIKEN, Wako, Saitama
  351-0198, Japan} 

\author{K. A.  Matveev} 

\altaffiliation{On leave from Duke University, Durham, NC 27708-0305.}
\affiliation{Materials Science Division, Argonne National Laboratory, 
Argonne, IL 60439, USA}

\date{\today}

\begin{abstract}
  We study the renormalization of a single impurity potential in
  one-dimensional interacting electron systems in the presence of magnetic
  field.  Using the bosonization technique and Bethe ansatz solutions, we
  determine the renormalization group flow diagram for the amplitudes of
  scattering of up- and down-spin electrons by the impurity in a quantum
  wire at low electron density and in the Hubbard model at less than 
  half filling.  In the absence of magnetic field the
  repulsive interactions are known to enhance backscattering and make the
  impurity potential impenetrable in the low-energy limit.  On the
  contrary, we show that in a strong magnetic field the interaction may
  suppress the backscattering of majority-spin electrons by the impurity
  potential in the vicinity of the weak-potential fixed point.  This
  implies that in a certain temperature range the impurity becomes almost
  transparent for the majority-spin electrons while it is impenetrable for
  the minority-spin ones.  The impurity potential can thus have a strong
  spin-filtering effect.
\end{abstract}

\pacs{
71.10.Pm, 
73.63.Nm, 
71.10.Fd 
}
\maketitle

\section{Introduction}
\label{I}

Quantum transport in one-dimensional (1D) electron systems has been a
subject of great interest for many years.  In one dimension the interplay
between electron-electron interaction and residual disorder is a crucial
factor determining transport properties.  It is well known that the
repulsive interactions between electrons strongly enhance the
backscattering of electrons by impurities at low
temperature.\cite{KaneF1992,FurusakiN1993} This phenomenon is a
manifestation of the fact that 1D electron systems at low temperatures
become Tomonaga-Luttinger (TL) liquids, with properties very different
from those of conventional Fermi liquids.  The renormalizations of the
electron scattering by impurities have been observed experimentally by
measuring temperature or bias dependence of the conductance of quantum
wires\cite{TaruchaHS1995,yacoby} and carbon nanotubes.\cite{bockrath,yao}

In this paper we explore the TL-liquid renormalizations of the potential
of a single impurity in the presence of a strong magnetic field $B$.  Such
a field causes significant polarization of electron spins.  This
polarization modifies the low-energy properties of the TL liquid,
resulting in qualitatively different renormalization group (RG) flows of
electron backscattering by the impurity.
In particular, when the energy band is less (more) than half filled,
repulsive interactions may \emph{decrease\/} the backscattering of
majority-spin electrons (holes) in the vicinity of the weak-impurity
fixed point, while increasing that of minority-spin electrons (holes).
This can be thought of as
enhancement of a spin-filtering effect due to interactions.

Renormalizations of the electron scattering by an impurity at zero
magnetic field has been studied analytically in the limiting cases of weak
and strong scatterer.\cite{KaneF1992,FurusakiN1993} The magnitude of the
impurity scattering of spin-$\lambda$ electrons ($\lambda=\upa,\dwa$)
is characterized by 
a small backscattering amplitude $v_\lambda$ in the former limit, 
and by a small amplitude $t_\lambda$ of tunneling through the impurity 
potential in the latter one.  The interactions between
electrons give rise to power-law renormalizations of these amplitudes at
low temperatures 
\be v_\lambda \propto
\left(\frac{T}{D}\right)^{\alpha_\lambda}, \qquad t_\lambda \propto
\left(\frac{T}{D}\right)^{\beta_\lambda},
\label{eq:renorm}
\ee 
where $T$ is the 
temperature and $D$ is the bandwidth.  The exponents $\alpha_\lambda$ and
$\beta_\lambda$ at $B = 0$ are determined by the electron-electron
interactions.  The repulsive interactions result in
$\alpha_\uparrow=\alpha_\downarrow < 0$ and
$\beta_\uparrow=\beta_\downarrow > 0$.  Thus, as the temperature $T$ is
lowered, the scattering of electrons by an impurity is enhanced in both
the weak and strong impurity limits.\cite{KaneF1992,FurusakiN1993}

Quantitative results\cite{KaneF1992,FurusakiN1993} for the exponents
$\alpha_\lambda$ and $\beta_\lambda$ have been obtained by using the
bosonization technique.  This method provides a convenient description of
the low-energy properties of 1D electron systems in terms of bosonic
fields $\phi_\lambda(x)$ and $\Pi_\lambda(x)$ satisfying the commutation
relations 
\be [\phi_\lambda(x), \Pi_{\lambda'}(x')]
=i\delta_{\lambda\lambda'}\delta(x-x').
\label{eq:phi-Pi-comm}
\ee
The effective Hamiltonian at $B = 0$ has
the spin-charge separated form 
$H = H_\rho + H_\sigma$ with 
\begin{equation}
H_\rho = \int \mathcal{H}_\rho dx,\quad
\mathcal{H}_\rho
=\frac{\hbar u_\rho}{2\pi}\left[\pi^2 K_\rho \Pi_\rho^2 
          +\frac{1}{K_\rho}(\partial_x\phi_\rho)^2\right],
\label{eq:H_rho}
\end{equation}
\begin{equation}
H_\sigma = \int \mathcal{H}_\sigma dx,\quad
\mathcal{H}_\sigma
=\frac{\hbar u_\sigma}{2\pi}
          \left[\pi^2 K_\sigma \Pi_\sigma^2 
          +\frac{1}{K_\sigma}(\partial_x\phi_\sigma)^2\right].
\label{eq:H_sigma}
\end{equation}
Here the fields $\phi_{\rho,\sigma}=(\phi_\uparrow \pm
\phi_\downarrow)/\sqrt2$ and $\Pi_{\rho,\sigma}=(\Pi_\uparrow \pm
\Pi_\downarrow)/\sqrt2$ describe excitations of the charge and spin modes,
$u_\rho$ and $u_\sigma$ are the velocities of the charge and spin
excitations, and $K_\rho$ and $K_\sigma$ are the TL-liquid parameters.  It
is known that for repulsive interactions $K_\rho$ is smaller than 1, while
$K_\sigma$ scales to 1 at low energies as required by the SU(2) symmetry
of the problem.\cite{SchulzCP2000,Giamarchi2003}

The results (\ref{eq:renorm}) for the renormalizations of the impurity
potential are obtained by adding to the Hamiltonian the perturbations
describing the impurity scattering and tunneling processes, and then
studying their scaling dimensions.  At $B = 0$ the exponents are related
to the TL-liquid parameters as
\be
\alpha_\lambda = \frac{K_\rho+K_\sigma}{2} - 1, 
\quad
\beta_\lambda = \frac12\left(\frac{1}{K_\rho}+\frac{1}{K_\sigma}\right) - 1.
\label{eq:exponents}
\ee
Hence, at $K_\rho < 1$ and $K_\sigma = 1$, 
the exponents $\alpha_\lambda$ is negative 
while $\beta_\lambda$ is positive.
It is interesting, however, that if the condition $K_\sigma=1$ 
imposed by the SU(2) symmetry of the problem is relaxed, 
the sign of the exponents $\alpha_\lambda$ and $\beta_\lambda$ 
may change depending on the value of $K_\rho < 1$.
For example, for a system with spin anisotropy resulting in 
$K_\sigma > 1$ and $2-K_\sigma<K_\rho<K_\sigma/(2K_\sigma-1)$, 
one has $\alpha_\lambda > 0$ and 
$\beta_\lambda > 0$.\cite{FurusakiN1993}$^,$\cite{KaneF1992,WongA1994} 
In this case, the potential of a weak impurity becomes 
an irrelevant perturbation, and the strength of the scatterer scales to
zero at low temperatures.
On the other hand, a scatterer with strength exceeding a certain 
critical value grows at $T\to0$, as indicated by positive exponent 
$\beta_\lambda$.

This analysis suggests that the renormalizations of impurity potential may
lead to a suppression of a weak impurity when a magnetic field is applied
to break the SU(2) symmetry. However, the effect of
magnetic field cannot be correctly described by using the Hamiltonian
(\ref{eq:H_rho}) and (\ref{eq:H_sigma}) since the spin-charge separation in
the 1D system is destroyed by the magnetic field.\cite{FrahmK1990,
  FrahmK1991} Thus, in order to understand the dependence of the exponents
$\alpha_\lambda$ and $\beta_\lambda$ on magnetic field, it
is necessary to generalize the low-energy Hamiltonian to the case $B \neq
0$.  Such a generalization was accomplished in Ref.\ 
\onlinecite{PencS1993} where the effective Hamiltonian for weakly
interacting electrons was obtained as a Gaussian model consisting of two
independent branches of bosonic excitations.  Using the effective
Hamiltonian, one can investigate the scaling of both the weak
backscattering due to an impurity, and weak tunneling through the
impurity potential.  We will see that if the magnetic field is
sufficiently strong, a spin-filtering phenomenon, in which the impurity
blocks the transport of minority-spin charge carriers and only weakly
scatters the majority-spin ones, can be realized in a certain regime
of the RG flow.

The paper is organized as follows.  We discuss the bosonization approach
to 1D interacting electron systems in a magnetic field in Sec.~\ref{II}\@.
The general form of the effective Hamiltonian is presented in
Sec.~\ref{IIA}\@.  
In Sec.~\ref{IIB}, we obtain the parameters in the effective Hamiltonian
for the 1D Hubbard model from their Bethe ansatz integral equations.  
In Sec.~\ref{IIC}, we derive the effective Hamiltonian for a
model describing quantum wires at low electron density.  In Sec.~\ref{III}
we discuss the renormalization of impurity potential $v_\lambda$.  The
scaling dimensions of the impurity potential and tunneling operators are
calculated in four limiting cases in Sec.~\ref{IIIA}\@.  The RG flow
diagrams are discussed in Sec.~\ref{IIIB}\@.  Section~\ref{IV} is devoted to
summary.

\section{effective Hamiltonian}
\label{II}
\subsection{Bosonization approach}
\label{IIA}

In this section we review the low-energy effective theory of 1D
interacting electrons in a magnetic field, following and extending the
bosonization approach introduced by Penc and S\'olyom.\cite{PencS1993}
This will be the basis of our analysis in the following sections.

To construct the low-energy theory, we first take
the noninteracting part of the Hamiltonian
and linearize the dispersions around the Fermi points $\pm k_{F \lambda}$.
The electron-field operators are expressed in terms of chiral fields as 
\be
\Psi_\lambda(x) = e^{ i k_{F \lambda} x} \Psi_{R \lambda}(x)
                + e^{-i k_{F \lambda} x} \Psi_{L \lambda}(x),
\label{eq:Psi}
\ee
where $\Psi_{R \lambda}$ ($\Psi_{L \lambda}$) is the field of 
right- (left-) moving electrons of spin $\lambda$.
Following the standard procedure,\cite{SchulzCP2000,Giamarchi2003} 
the chiral fields are bosonized as 
\be
\Psi_{P \lambda}(x) = \frac{\kappa_\lambda}{\sqrt{2\pi a}}
e^{i s_P \varphi_{P \lambda}(x)},
\ee
where
\be
s_P = 
\begin{cases}
+ & \mbox{for $P = R$},\cr
- & \mbox{for $P = L$},\cr
\end{cases}
\ee
$\kappa_\lambda$
is the Klein factor satisfying 
$\{ \kappa_\lambda, \kappa_{\lambda'} \} = 2\delta_{\lambda \lambda'}$
and $\kappa_\lambda^\dagger=\kappa_\lambda^{}$,
and $a$ is a short-distance cutoff.
The bosonic fields obey  commutation relations
\bea
&&
\left[ \varphi_{P \lambda}(x), \varphi_{P \lambda'}(y) \right] 
= i \pi s_P \delta_{\lambda \lambda'} {\rm sgn}(x-y),
\nn \\
&&
\left[ \varphi_{R \lambda}(x), \varphi_{L \lambda'}(y) \right] 
= - i \pi \delta_{\lambda \lambda'}.
\label{eq:varphicomm}
\eea
The Hamiltonian density of noninteracting electrons is given in terms
of the chiral fields by
\be
\mathcal{H}_0
= \frac{\hbar}{4\pi} \sum_\lambda u_\lambda
\left[ (\partial_x\varphi_{R \lambda})^2 
     + (\partial_x\varphi_{L \lambda})^2 \right],
\label{eq:Ham0}
\ee
where $u_\lambda > 0$ are the velocities of 
the linearized dispersion of the spin-$\lambda$ branch.

The interactions between electrons result in two-particle scattering
processes.  In the most general case, they can be classified\cite{solyom}
into the following four types: backward scattering (the $g_1$ process),
forward scattering ($g_2$), Umklapp scattering ($g_3$), and scattering
within one branch ($g_4$).  Among these scattering processes, the $g_1$
interaction between electrons with opposite spins can be discarded since
$k_{F\uparrow}\ne k_{F\downarrow}$.  The $g_1$ interaction between
electrons with equal spins is equivalent to the $g_2$ scattering.
Furthermore, since the electron density is assumed to be incommensurate
with the lattice, the Umklapp ($g_3$) scattering can be ignored.  As a
result, the most general form of the quadratic part of the interaction
Hamiltonian density reads 
\bea \mathcal{H}_{\mathrm{int}} &=&
\frac{\hbar}{8\pi^2} \sum_{\lambda, P} \left[ g_{2\lambda}
  \frac{d\varphi_{P \lambda}}{dx} \frac{d\varphi_{\bar{P} \lambda}}{dx}
+ g_{2 \perp} 
\frac{d\varphi_{P \lambda}}{dx} \frac{d\varphi_{\bar{P} \bar{\lambda}}}{dx}
\right.\nn \\
&& \hspace*{14mm}\left.
+ g_{4\lambda}
\frac{d\varphi_{P \lambda}}{dx} \frac{d\varphi_{P \lambda}}{dx}
+ g_{4 \perp} 
\frac{d\varphi_{P \lambda}}{dx} \frac{d\varphi_{P \bar{\lambda}}}{dx}
\right],
\nn \\
\label{eq:Hamint}
\eea
where the coupling constants $g_{i\lambda,\perp}$ are real, and
\be
\bar\lambda=
\begin{cases}
\downarrow & \mbox{for $\lambda=\uparrow$},\cr
\uparrow & \mbox{for $\lambda=\downarrow$},\cr
\end{cases}
\qquad
\bar P=
\begin{cases}
L & \mbox{for $P=R$},\cr
R & \mbox{for $P=L$}.\cr
\end{cases}
\ee
Combining Eqs.\ (\ref{eq:Ham0}) and (\ref{eq:Hamint}), 
we find the total effective Hamiltonian density written
in the matrix form
\be
\widetilde{\mathcal{H}} = 
\mathcal{H}_0 + \mathcal{H}_{\rm int}
= \frac{\hbar}{4\pi} \partial_x {\bm \varphi}^T(x)~
\hat{\mathcal{H}} ~\partial_x {\bm \varphi}(x),
\label{eq:Hameff}
\ee
where 
${\bm \varphi}^T = (\varphi_{R \uparrow}, \varphi_{L \uparrow}, 
\varphi_{R \downarrow}, \varphi_{L \downarrow})$,
\be
\hat{\mathcal{H}} =
\left(\begin{array}{cccc}
u_\uparrow + \tilde{g}_{4\uparrow} & \tilde{g}_{2 \uparrow}
& \tilde{g}_{4 \perp} & \tilde{g}_{2 \perp}
\\
\tilde{g}_{2 \uparrow} & u_\uparrow + \tilde{g}_{4\uparrow}
& \tilde{g}_{2 \perp} & \tilde{g}_{4 \perp} 
\\
\tilde{g}_{4 \perp} & \tilde{g}_{2 \perp} 
& u_\downarrow + \tilde{g}_{4\downarrow} & \tilde{g}_{2 \downarrow}
\\
\tilde{g}_{2 \perp} & \tilde{g}_{4 \perp}
& \tilde{g}_{2 \downarrow} & u_\downarrow + \tilde{g}_{4\downarrow}
\end{array}
\right),
\label{eq:14}
\ee
and $\tilde{g}_{i\lambda,\perp} = g_{i\lambda,\perp}/(2\pi)$.

We show in Appendix~\ref{A} that the matrix $\hat{\mathcal{H}}$ can be
brought to the form
\be
\hat{\mathcal{H}} = \sum_{P,\nu} u_\nu 
                    {\bm \omega}_{P\nu} ({\bm \omega}_{P\nu})^T,
\label{eq:diagH}
\ee 
with real vectors ${\bm \omega}_{P\nu}$ satisfying the orthonormal
conditions
\
\begin{equation}
  \label{eq:P-orthogonal}
  ({\bm \omega}_{P\nu})^T \hat{C} {\bm \omega}_{P'\nu'} 
    = s_P \delta_{PP'}\delta_{\nu\nu'}.
\end{equation}
Here the subscript $\nu$ takes two possible values, which we will
denote as $c$ and $s$, parameters $u_c$ and $u_s$ are positive,
and the matrix $\hat{C}$ is defined as $\hat{C} = {\rm diag}(1, -1, 1, -1)$
and accounts for the sign factor $s_P$ in the
commutation relations (\ref{eq:varphicomm}).

We then introduce chiral fields 
\be
\tilde{\varphi}_{P\nu}(x) = ({\bm \omega}_{P\nu})^T {\bm \varphi}(x),
\label{eq:tphi}
\ee 
satisfying the same commutation relations as the original fields
$\varphi_{P\lambda}$, see Eq.~(\ref{eq:varphicomm}).  In terms of
these new fields the Hamiltonian (\ref{eq:Hameff}) takes the simple form
\be 
\widetilde{\mathcal{H}} = \frac{\hbar}{4\pi}
\sum_{\nu=c,s} u_\nu \{[\partial_x \tilde{\varphi}_{R\nu}(x)]^2
                      +[\partial_x \tilde{\varphi}_{L\nu}(x)]^2\}.
\ee
The positive constants $u_c$ and $u_s$ have the meanings of the
velocities of the two types of elementary excitations of the
Hamiltonian $\widetilde{\mathcal{H}}$.  We refer to these excitations
as the holons and spinons.  We then introduce the fields
\be
\tilde{\phi}_\nu 
= \frac{1}{2} (\tilde{\varphi}_{L \nu} + \tilde{\varphi}_{R \nu}),~~~
\widetilde{\Pi}_\nu = \frac{1}{2\pi} 
\partial_x (\tilde{\varphi}_{L \nu} - \tilde{\varphi}_{R \nu}),
\label{eq:tildephi}
\ee
and rewrite the effective Hamiltonian density
as
\be
\widetilde{\mathcal{H}} = \frac{\hbar}{2\pi} \sum_{\nu = c, s}
u_\nu  \left[ \pi^2 \widetilde{\Pi}_\nu^2 
+ \left( \partial_x \tilde{\phi}_\nu \right)^2 \right].
\label{eq:tHam}
\ee
Hence the system of 1D
interacting electrons in a magnetic field can be described as a
two-component TL liquid.\cite{PencS1993} We note that at $B = 0$ the holon
and spinon modes reduce to the charge and spin modes in Eqs.\ 
(\ref{eq:H_rho}) and (\ref{eq:H_sigma}), respectively.

The fields $\tilde{\phi}_\nu$ and $\widetilde{\Pi}_\nu$ are related to
the original bosonic fields 
\be
\phi_\lambda = \frac{1}{2} (\varphi_{L \lambda} + \varphi_{R \lambda}),
\quad
\Pi_\lambda = \frac{1}{2\pi} 
\partial_x (\varphi_{L \lambda} - \varphi_{R \lambda})
\label{eq:phiPi}
\ee
used in the Hamiltonian (\ref{eq:H_rho}), (\ref{eq:H_sigma})  
through Eqs.\ (\ref{eq:tphi}) and (\ref{eq:tildephi}).
Due to the parity symmetry of the system,
this linear relation is simplified to
\be
\left(\begin{array}{c}
\phi_\upa
\\
\phi_\dwa
\end{array}
\right)
= \hat{A}^T
\left(\begin{array}{c}
\tilde{\phi}_c
\\
\tilde{\phi}_s
\end{array}
\right)
,~~~
\left(\begin{array}{c}
\Pi_\upa
\\
\Pi_\dwa
\end{array}
\right)
= \hat{A}^{-1}
\left(\begin{array}{c}
\widetilde{\Pi}_c
\\
\widetilde{\Pi}_s
\end{array}
\right),
\label{eq:tphi-to-phi}
\ee
where the real matrix $\hat{A}$ can be obtained from the vectors
${\bm \omega}_{P\nu}$, see Appendix~\ref{B}.
Note that our Hamiltonian (\ref{eq:tHam}) is determined by 6 parameters:
the velocities of holons and spinons $u_c$ and $u_s$, as well as the four
elements of the transformation matrix $\hat A$.  

An alternative approach 
to bosonized description of 1D systems in magnetic field was used in
Ref.~\onlinecite{KimuraKAL1994}.  In their theory the Hamiltonian depends
on 5 parameters, namely the velocities $u_\rho$ and $u_\sigma$ of the
charge and spin modes in the absence of the field, the TL-liquid parameters
$K_\rho$ and $K_\sigma$, and the difference of velocities of
spin-$\uparrow$ and spin-$\downarrow$ electrons induced by the field.  (In
the realistic case of spin-independent interactions between electrons,
$K_\sigma=1$, and the number of parameters is further reduced to 4.)  This
simplification of the theory\cite{KimuraKAL1994} occurred because the
magnetic field dependence of the coupling constants describing the
electron-electron interactions was neglected.  We believe the approach of
Ref.~\onlinecite{KimuraKAL1994} is therefore inapplicable beyond the
regime of weak magnetic field.

\subsection{One-dimensional Hubbard model}
\label{IIB}

The effective Hamiltonian (\ref{eq:tHam}) depends on six parameters,
$u_c$, $u_s$, and the four matrix elements of $\hat{A}$.  In the case
of exactly solvable models, these parameters can be obtained exactly
by solving the integral equations of the Bethe
ansatz.\cite{FrahmK1990,FrahmK1991} Here we discuss one such case,
namely the 1D Hubbard model.

The original Hamiltonian of the 1D Hubbard model has the form
\bea
H_{\rm Hub} &=& -t \sum_{l,\lambda}
 (c^\dagger_{l,\lambda} c^{}_{l+1,\lambda} + {\rm H.c.})
+ U \sum_l n_{l,\upa} n_{l,\dwa}
\nn \\
&&- \frac{B}{2} \sum_l (n_{l,\upa}-n_{l,\dwa}),
\eea
where $c^\dagger_{l,\lambda}$ ($c_{l,\lambda}$) is the creation
(annihilation) operator of spin-$\lambda$ electron at the $l$th site,
$n_{l,\lambda} = c^\dagger_{l,\lambda} c^{}_{l,\lambda}$, the matrix
element $t$ accounts for hopping between neighboring lattice sites, and
$U$ is the strength of the on-site repulsion; $t,U>0$.  Throughout
this paper, we concentrate on the case of less than half-filling,
i.e., the total electron density $n < 1$.  We will comment on the case
of $n > 1$ in Sec.~\ref{IV}.

The 1D Hubbard model allows for exact solution by Bethe ansatz at
arbitrary value of the field $B$.  This solution enables one to obtain
not only the velocities $u_c$ and $u_s$, but also the asymptotics of
various correlation functions at large distances.  The latter are
expressed in terms of the so-called dressed charges $Z_{\nu
  \nu'}$, which can be found exactly by solving integral equations of
the Bethe ansatz.\cite{FrahmK1990,FrahmK1991} For example, the dressed
charge matrix is given by 
\be \left(\begin{array}{cc} Z_{cc} & Z_{cs}
    \\
    Z_{sc} & Z_{ss}
\end{array}
\right)
=
\left(\begin{array}{cc}
\xi & 0
\\
\frac{1}{2}\xi & 1/\sqrt{2}
\end{array}
\right)
\ee
at $B=0$, and 
\be
\left(\begin{array}{cc}
Z_{cc} & Z_{cs}
\\
Z_{sc} & Z_{ss}
\end{array}
\right)
=
\left(\begin{array}{cc}
1 & \frac{2}{\pi} \arctan\left[ \frac{4 t \sin(\pi n)}{U} \right]
\\
0 & 1
\end{array}
\right)
\ee
in the saturation limit where the electron spins are fully
polarized.
Here the dressed charge $\xi$ is defined in Eq.\ (5.1) in
Ref.\ \onlinecite{FrahmK1990} and takes values 
in the range $1 \le \xi \le \sqrt{2}$.
As the magnetization increases, the dressed charges change continuously 
between the values in the limiting cases.

By comparing the critical exponents obtained from the Bethe ansatz 
with those from the effective Hamiltonian (\ref{eq:tHam}),
one can relate the matrix $\hat{A}$ to the dressed charges 
$Z_{\nu \nu'}$ as,\cite{PencS1993,Cabra2001}
\be
\hat{A} =
\left(\begin{array}{cc}
A_{11} & A_{12}
\\
A_{21} & A_{22}
\end{array}
\right)
=
\left(\begin{array}{cc}
Z_{cc} - Z_{sc} & Z_{sc}
\\
Z_{ss} - Z_{cs} & -Z_{ss}
\end{array}
\right).
\label{eq:AforHub}
\ee
We will use these results in Sec.~\ref{III}.

As a simple model for an impurity potential in the Hubbard model,
we can take the on-site
potential
\begin{equation}
V_\mathrm{Hub}=V_0(n_{0,\uparrow}+n_{0,\downarrow}).
\end{equation}
To analyze low-energy transport properties, we take the continuum limit,
where the density operator is approximated as
$n_{0,\lambda}\approx\Psi^\dagger_\lambda(0)\Psi^{}_\lambda(0)$.
With the chiral electron fields the potential is further reduced to
the form $V=V_\uparrow+V_\downarrow$, where
\begin{equation}
V_\lambda=v_\lambda
\left[\Psi^\dagger_{R\lambda}(0)\Psi^{}_{L\lambda}(0)
+\Psi^\dagger_{L\lambda}(0)\Psi^{}_{R\lambda}(0)\right].
\label{V_lambda}
\end{equation}
Here we have kept only the backward scattering terms and
discarded the forward scattering ones, as the latter do not
affect the conductance.
We have also introduced spin-dependent backscattering amplitude
$v_\lambda$.
Finally, Eq.~(\ref{V_lambda}) is written in terms of the bosonic fields
as
\begin{equation}
V_\lambda 
= -\frac{v_\lambda}{\pi a} \cos[2\phi_\lambda(x=0)].
\label{eq:Vimp}
\end{equation}
We set the amplitudes $v_\lambda$ to be positive, which is always
possible by the transformation
$\phi_\lambda\to\phi_\lambda+\mathrm{const}$.

\subsection{Quantum wires at low electron density}
\label{IIC}

Here we derive the low-energy effective Hamiltonian of 
a quantum wire in the low-density limit,
where the effective Hamiltonian (\ref{eq:tHam}) takes
a particularly simple form similar to Eqs.\ (\ref{eq:H_rho})
and (\ref{eq:H_sigma}), as we will see below.

When the electron density in the wire is very low,
the electron-electron interactions are effectively
very strong.
In the limit of infinitely strong repulsion, electrons can never
occupy the same position in space
and can be viewed as distinguishable particles.
As a result, the energy of the electron system becomes independent of 
the electron spins.
At strong but finite interactions, the electrons in the wire
can exchange their positions, and the spins of neighboring electrons
are weakly coupled to each other.
The resulting spin dynamics is described by 
the Heisenberg model,
\be
H_\sigma = J \sum_l {\bm S}_{l}\cdot{\bm S}_{l+1}.
\label{eq:Heisenberg}
\ee
Hence the Hamiltonian of the wire at zero magnetic field $B = 0$ takes 
the spin-charge separated form\cite{matveev1,matveev2}
$H = H_\rho + H_\sigma$ with the two terms 
given by Eqs.~(\ref{eq:H_rho}) and (\ref{eq:Heisenberg}).

At energy scales below the exchange constant $J$ the Hamiltonian 
(\ref{eq:Heisenberg}) can be bosonized,\cite{SchulzCP2000,Giamarchi2003} 
and the form (\ref{eq:H_sigma}) of the Hamiltonian density
$\mathcal{H}_\sigma$ 
is recovered.
The advantage of using the Heisenberg form (\ref{eq:Heisenberg}) is that 
the magnetic field $B$ can be easily incorporated by adding a term 
$-|g|\mu_B B S^z$, where $g$ is Lande factor and $\mu_B$ is Bohr magneton.
The field $B$ polarizes the spins and results 
in finite magnetization.\cite{Griffiths1964}
In the following, it will be convenient to parametrize the Hamiltonian 
by a relative magnetization $m$ defined as 
$m=(n_\uparrow-n_\downarrow)/(n_\uparrow+n_\downarrow)$, 
where $n_{\uparrow,\downarrow}$ are the densities of electrons 
with given spin components.
At $m<1$ the Hamiltonian of the Heisenberg model in a magnetic field 
can be bosonized to the form (\ref{eq:H_sigma}), 
with the velocity $u_\sigma$ and the coupling parameter $K_\sigma$ 
becoming functions of $m$.\cite{Haldane1980} 
As $m$ varies from 0 to 1, the velocity $u_\sigma(m)$ changes 
from $\pi J/[2\hbar(n_\uparrow+n_\downarrow)]$ to zero,
and $K_\sigma(m)$ grows from 1 to 2.

Using the separation of charge and spin variables in the form
(\ref{eq:H_rho}), (\ref{eq:Heisenberg}) and above mentioned properties of
the Heisenberg model, one can conclude that the low-energy Hamiltonian
density of strongly interacting electron system in a magnetic field has
the form 
\bea \widetilde{\mathcal{H}} &=&
\frac{\hbar u_\rho}{2}\left[ 
    \pi K_\rho (\Pi_\rho + m\Pi_\sigma)^2 +\frac{1}{\pi
    K_\rho}(\partial_x\phi_\rho)^2 \right]
\nonumber\\
&&+ \frac{\hbar u_\sigma(m)}{2}\left[\pi K_\sigma(m) \Pi_\sigma^2
  +\frac{[\partial_x(\phi_\sigma-m\phi_\rho)]^2}{\pi K_\sigma(m)} \right].
\nn \\
&&
\label{eq:Hamiltonian_magnetized}
\eea
The first line of the Hamiltonian (\ref{eq:Hamiltonian_magnetized}) 
describes the charge density excitations (holons) of the electron system.
Since the coupling of the spins is very weak, the magnetic field 
polarizing the spins does not affect the speed of holons $u_\rho$.
The form of the holon part is thus essentially equivalent to 
Eq.\ (\ref{eq:H_rho}), with the addition of the term $m\Pi_\sigma$ 
to the momentum density.
This correction does not affect the dynamics of the holons, 
as $[\Pi_\sigma,\partial_x\phi_\rho]=0$.
On the other hand, the addition of $m\Pi_\sigma$ to the momentum density 
ensures that the holon wave carries the spin current due to 
the finite magnetization $m$ of the ground state.
Indeed, the equation of motion for the holon wave results in the relation 
$\dot\phi_\sigma=m\dot\phi_\rho$ between the spin and charge currents.

The form of the spinon part of the Hamiltonian essentially reproduces the
bosonized Hamiltonian of the Heisenberg model at finite magnetization; in
particular, the dependences $u_\sigma(m)$ and $K_\sigma(m)$ are equivalent
to those discussed in Ref.~\onlinecite{Haldane1980}.  The only difference
is the addition of the term $-m\,\partial_x\phi_\rho$ to the spin density.
Due to the commutation relation $[\phi_\rho, \Pi_\sigma]=0$, this term
does not change the spin dynamics.  However, its presence ensures that in
the spinon ground state the spin and charge densities are proportional to
each other: $\partial_x\phi_\sigma = m \partial_x\phi_\rho$.

The effective Hamiltonian (\ref{eq:Hamiltonian_magnetized}) can be easily
brought to the standard form (\ref{eq:tHam}), with the matrix $\hat A$
taking the form
\be
\hat{A} =
\left(\begin{array}{cc}
A_{11} & A_{12}
\\
A_{21} & A_{22}
\end{array}
\right)
=
\left(\begin{array}{cc}
\sqrt{\frac{K_\rho}{2}}(1+m) & \sqrt{\frac{K_\rho}{2}}(1-m)
\\
\sqrt{\frac{K_\sigma(m)}{2}} & - \sqrt{\frac{K_\sigma(m)}{2}}
\end{array}
\right).
\label{eq:AforQW}
\ee 
In general, the parameter $K_\rho$ is non-universal.  In the limit of
strong short-range interaction it can be deduced from the well known
properties of the Hubbard model, and one finds $K_\rho=\frac12$.  For
longer range interactions one expects $K_\rho<\frac12$.  On the other
hand, the parameter $K_\sigma(m)$ is the TL-liquid parameter for the
Heisenberg spin chain in magnetic field, which can be determined exactly
by solving the Bethe ansatz integral
equations.\cite{Haldane1980,KorepinBI1993} These results will be used in
Sec.~\ref{III} to investigate the renormalizations of impurity potential.

Our discussion in this section assumed arbitrary range of interactions
between the electrons in the quantum wire.  In experiments the range of
the Coulomb repulsion between electrons is usually longer than the
distance between particles.  However, the range of the interactions is
limited by the presence of metal gates in the vicinity of the wire. One
can show that at electron densities below $a_B/d^2$ the range of the
interactions is short compared with the distance between the
electrons.\cite{matveev2} [Here $d$ is the distance from the wire to the
nearest gate, and $a_B$ is the effective Bohr radius in the material;
$a_B\approx 10$\,nm in GaAs.]  In this special case the electrons in the
quantum wire can be described\cite{matveev2} by the Hubbard model in the
limit of low filling, $n\to0$, when the discreteness of the lattice can be
neglected.  In particular, the fact that the spin excitations are those of
the Heisenberg model (\ref{eq:Heisenberg}) corresponds to the well-known
property\cite{ogata} of the Hubbard model in the limit $U/t\to\infty$.  In
this limit the parameter $K_\rho$ takes the value $1/2$.  We have checked
that at $K_\rho=1/2$ our result (\ref{eq:AforQW}) for the matrix $\hat A$
coincides with the result (\ref{eq:AforHub}) for the Hubbard model with the
dressed charges $Z_{\nu\nu'}$ found in the limit $U/t\to\infty$ in
Ref.~\onlinecite{sugiyama}.

Similarly to the case of the Hubbard model, the presence of an impurity in
a quantum wire results in backscattering of electrons.  We will use 
expressions (\ref{V_lambda}) and (\ref{eq:Vimp}) for the 
perturbations in the Hamiltonian,
cf.~Refs.~\onlinecite{KaneF1992,FurusakiN1993}.

\section{Renormalization-group analysis}
\label{III}

In this section we discuss the renormalizations of 
impurity backscattering amplitude $v_\lambda$
and tunneling amplitude $t_\lambda$ using
the effective Hamiltonian obtained in Sec.~\ref{II}.
We will consider the following four limiting cases:
\begin{subequations}
\label{eq:cases}
\begin{eqnarray}
&& \mbox{$|v_\uparrow|\ll D$ and $|v_\downarrow|\ll D$},
\label{eq:a}\\
&& \mbox{$|t_\uparrow|\ll D$ and $|t_\downarrow|\ll D$},
\label{eq:b}\\
&& \mbox{$|v_\uparrow|\ll D$ and $|t_\downarrow|\ll D$},
\label{eq:c}\\
&& \mbox{$|t_\uparrow|\ll D$ and $|v_\downarrow|\ll D$}.
\label{eq:d}
\end{eqnarray}
\end{subequations}
Scaling dimensions of the backscattering and tunneling operators in the
four cases are summarized in Table~\ref{tab:sdim}.  Evaluating the scaling
dimensions quantitatively, we construct the RG flow diagram in the
$v_\upa$-$v_\dwa$ plane.  We will see that the RG flow diagram changes
drastically when a sufficiently strong field is applied.

\begin{table}
\caption{
\label{tab:sdim}
Scaling dimensions of the backscattering and tunneling operators 
$V_\lambda$ and $T_\lambda$ in the four limiting cases 
(\ref{eq:a})--(\ref{eq:d}).
The expressions of $x_{V \lambda}$ and $x_{T \lambda}$ are 
given in Eqs.\ (\ref{eq:xV}) and (\ref{eq:xT}), respectively.
}
\begin{ruledtabular}
\begin{tabular}{ccccc}
limits & $V_\upa$ & $V_\dwa$ & $T_\upa$ & $T_\dwa$ \\
\hline
(a) $|v_\uparrow|, |v_\downarrow|\ll D$ & $x_{V \upa}$ & $x_{V \dwa}$ & --- & --- \\
(b) $|t_\uparrow|, |t_\downarrow|\ll D$ & --- & --- & $x_{T \upa}$ & $x_{T \dwa}$ \\
(c) $|v_\uparrow|, |t_\downarrow|\ll D$ & $1/x_{T \upa}$ & --- & --- & $1/x_{V \dwa}$ \\
(d) $|t_\uparrow|, |v_\downarrow|\ll D$ & --- & $1/x_{T \dwa}$ & $1/x_{V \upa}$ & --- \\
\end{tabular}
\end{ruledtabular}
\end{table}

\subsection{Scaling dimensions of backscattering and tunneling operators}
\label{IIIA}
\subsubsection{Weak-potential limit $|v_\upa|, |v_\dwa| \ll D$}
\label{IIIA1}
First, we discuss the renormalization of the impurity scattering 
in the limit where both $v_\upa$ and $v_\dwa$ are weak, 
$|v_\upa|, |v_\dwa| \ll D$.
To find the scaling dimension of the operator (\ref{eq:Vimp}), we calculate 
the ground-state correlation function of $V_\lambda$ 
using the action for the pure system (\ref{eq:tHam}) given by
\bea
S &=&
\sum_{\nu = c, s} \int_0^\beta d\tau \int^\infty_{-\infty} dx \left[
\frac{1}{2\pi u_\nu} \left( \partial_\tau \tilde{\phi}_\nu(x, \tau) \right)^2
\right.
\nn \\
&&\left.~~~~~~~~~~~~~~~~~~~~~~~~~
+ \frac{u_\nu}{2\pi} \left( \partial_x \tilde{\phi}_\nu(x, \tau) \right)^2
\right],
\label{eq:action}
\eea
where $\beta = 1/T$ and $\tau$ is the imaginary time.
Since $V_\lambda$ depends only on
$\phi^0_\lambda(\tau) \equiv \phi_\lambda(x = 0, \tau)$,
we integrate out the fields $\tilde\phi_\nu(x,\tau)$ at $x \ne 0$ 
to derive the effective action for $\tilde\phi_\nu(x=0,\tau)$,
\be
S_0 = \sum_{\omega_n} \frac{|\omega_n|}{\pi} \left[
\left|\tilde{\phi}_c^0(\omega_n)\right|^2
 + \left|\tilde{\phi}_s^0(\omega_n)\right|^2
\right].
\label{eq:Seff}
\ee
Here we have introduced the Fourier transform
\be
\tilde{\phi}_\nu^0(\omega_n) 
= \frac{1}{\sqrt{\beta}}\int^\beta_0 e^{i\omega_n \tau}
\tilde\phi_\nu(x=0,\tau) d\tau
\ee
with $\omega_n = 2\pi n/\beta$.
Using the effective action, the imaginary-time
correlation function is calculated as
\bea
&&
\langle e^{2i\phi_\lambda^0(\tau)} e^{-2i\phi_\lambda^0(0)} \rangle
\nn \\
&&~~~= \frac{1}{Z} \int \mathcal{D}\tilde\phi_c^0
 \mathcal{D}\tilde\phi_s^0 
e^{-S_0+2i[\phi_\lambda^0(\tau)-\phi_\lambda^0(0)]},
\label{eq:Vlambdacor}
\eea
where 
\be
Z = \int \mathcal{D}\tilde\phi_c^0 \mathcal{D}\tilde\phi_s^0 e^{-S_0}
\label{eq:partition}
\ee 
is the partition function.
A straightforward calculation of Eq.\ (\ref{eq:Vlambdacor}) with 
Eqs.\ (\ref{eq:Seff}) and (\ref{eq:tphi-to-phi}) gives
the correlation functions in the limit $\beta\to\infty$,
\be
\langle V_\lambda(\tau) V_\lambda(0) \rangle
\propto \tau^{-2x_{V\lambda}},
\ee
with the scaling dimensions  $x_{V\lambda}$ given by
\begin{subequations}
\begin{eqnarray}
x_{V \upa}& = & A_{11}^2 + A_{21}^2,
\label{eq:xVup}
\\
x_{V \dwa}& = & A_{12}^2 + A_{22}^2.
\end{eqnarray}
\label{eq:xV}
\end{subequations}
The exponents $\alpha_\lambda$ 
in Eq.\ (\ref{eq:renorm}) in this limit are given by
\be
\alpha_{\lambda}^{(a)} = x_{V \lambda} - 1.
\ee

\subsubsection{Weak-tunneling limit $|t_\upa|, |t_\dwa| \ll D$}
\label{IIIA2}
The renormalizations of tunneling through a strong impurity potential were
studied using several different
approaches.\cite{KaneF1992,FurusakiN1993,FurusakiM1995} The discussion
given below uses the method of Ref.~\onlinecite{FurusakiM1995}.

Let us consider the tunneling of a spin-$\upa$ electron 
through the impurity potential at $x = 0$. 
Since the potential amplitudes $v_\lambda$ ($\lambda=\upa,\dwa$)
are assumed to be very large,
the fields $\phi_\lambda^0$ are pinned at the minima of 
the potential $V_\lambda$ [Eq.\ (\ref{eq:Vimp})], 
i.e., $\phi_\lambda^0 = \pi l_\lambda$,
where $l_\lambda$ are integers.
The tunneling of a spin-$\upa$ electron through the potential barrier
is equivalent to a sudden change of $\phi_\upa^0$ between
neighboring minima, say, from $\phi^0_\upa = 0$
to $\phi^0_\upa = \pi$, i.e., a jump in $\phi_\upa^0$ by $\pi$.
Let us denote the operator for this tunneling process by $T_\upa$.
The correlation function
$\langle T^\dagger_\upa(\tau)T^{}_\upa(0)\rangle$ is then
obtained from
\be
\langle T^\dagger_\upa(\tau)T^{}_\upa(0)\rangle
\propto\exp(-S_0)\Big|_{\phi_\upa^0(\tau')=\pi\theta(\tau'-\tau),
                                                    \phi_\dwa^0=0},
\ee
where we have ignored small fluctuations of $\phi_\lambda^0$ around
the potential minima. [Here $\theta(\tau)$ is the unit step function.]
From Eq.\ (\ref{eq:tphi-to-phi}) we substitute
\be
\left(
\begin{array}{c}
\tilde\phi_c^0(\omega_n) \\
\tilde\phi_s^0(\omega_n)
\end{array}
\right)
=(\hat{A}^T)^{-1}
\left(
\begin{array}{c}
i\pi(1-e^{i\omega_n\tau})/\omega_n\sqrt{\beta} \\
0
\end{array}
\right)
\ee
into $\exp(-S_0)$ to find
\be
\langle T^\dagger_\upa(\tau)T^{}_\upa(0)\rangle
\propto\tau^{-2x_{T\uparrow}},
\ee
where the scaling dimension is
\begin{subequations}
\label{eq:xT}
\be
x_{T \upa} = [(A^{-1})_{11}]^2 + [(A^{-1})_{12}]^2
           = \frac{A_{22}^2 + A_{12}^2}{(\det \hat{A})^2}.
\label{eq:xTup}
\ee
Similarly we have
\be
x_{T \dwa} = [(A^{-1})_{21}]^2 + [(A^{-1})_{22}]^2
           = \frac{A_{21}^2 + A_{11}^2}{(\det \hat{A})^2}.
\ee
\end{subequations}
The exponents for the tunneling amplitude $t_\lambda$ 
in Eq.\ (\ref{eq:renorm}) in this limit are
\be
\beta_\lambda^{\rm (b)} = x_{T \lambda} - 1.
\ee

\subsubsection{Asymmetric limits 
$|v_\upa|, |t_\dwa| \ll D$ and $|t_\upa|, |v_\dwa| \ll D$}
\label{IIIA3}
Next we consider the asymmetric limit where the potential scattering is
weak for the spin-$\upa$ electrons but strong for the spin-$\dwa$ electrons.
Although such an extremely spin-selective scattering is not likely 
to be realized with a bare impurity potential, we will see in
Sec.~\ref{IIIB} that this is indeed realized for some RG trajectory
if a magnetic field applied is sufficiently large.

The scaling dimension of the potential $V_\upa$ in the limit 
$|v_\upa|, |t_\dwa| \ll D$ can be found in a similar way 
to the weak potential limit discussed in Sec.~\ref{IIIA1}.
The only difference is that in the present case 
$\phi_\dwa(x=0)$ is pinned at a potential minimum $\pi l_\dwa$ by 
the strong impurity potential $V_\dwa$.
The asymptotic form of the ground-state correlation function is 
then obtained as 
\bea
\langle V_\upa(\tau) V_\upa(0) \rangle
&\propto&
\left.
\int\!\mathcal{D}\phi_\upa^0 \,
e^{-S_0+2i[\phi_\upa^0(\tau)-\phi_\upa^0(0)]}
\right|_{\phi_\dwa^0=0}
\nonumber\\
&\propto&\tau^{-2/x_{T\uparrow}},
\eea
with $x_{T\uparrow}$ given by Eq.~(\ref{eq:xTup}).
Therefore the scaling dimension of $V_\upa$ in this limit 
is $1/x_{T \upa}$,
and the exponent $\alpha_\upa$ is given by
\be
\alpha_{\upa}^{\rm (c)} = \frac{1}{x_{T \upa}} - 1.
\ee
Similarly, the scaling dimension of $V_\dwa$ 
in the limit $|t_\upa|, |v_\dwa| \ll D$ is found to be $1/x_{T \dwa}$, 
resulting in the exponent
$\alpha_{\dwa}^{\rm (d)} = 1/x_{T \dwa} - 1$.

The scaling of the tunneling operator $T_\upa$ in the limit 
$|t_\upa|, |v_\dwa| \ll D$ can be studied in a similar manner
to that in Sec.\ III A2.
In the present case, however, the potential $V_\dwa$ is weak,
and the field $\phi_\dwa^0$ can fluctuate almost freely.
Therefore, to find the scaling dimension in lowest order in $v_\dwa$,
we first integrate out $\phi_\dwa^0$  in $S_0$
to obtain the effective action for $\phi_\upa^0$, into which
we substitute $\phi_\upa^0(\tau')=\pi\theta(\tau'-\tau)$.
This yields
\bea
\langle T_\upa^\dagger(\tau)T_\upa^{}(0)\rangle
&\propto&
\left.
\int\mathcal{D}\phi_\dwa^0 e^{-S_0}
\right|_{\phi_\upa^0(\tau')=\pi\theta(\tau'-\tau)}
\nonumber\\
&\propto&\tau^{-2/x_{V\uparrow}},
\eea
with $x_{V\uparrow}$ given by Eq.~(\ref{eq:xVup}).  We thus conclude that
the scaling dimension of $T_\upa$ in the limit (\ref{eq:d}) is $1/x_{V
  \upa}$, and the exponent $\beta_{\upa}^{\rm (d)} = 1/x_{V \upa} - 1$.
Similarly, the scaling dimension of $T_\dwa$ in the limit (\ref{eq:c}) is
$1/x_{V \dwa}$, and the exponent $\beta_{\dwa}^{\rm (c)} = 1/x_{V \dwa} - 1$.

\subsection{RG flow diagram}
\label{IIIB}
In the preceding sections we have found that the scaling dimensions of the
backscattering and tunneling operators in the four limits
(\ref{eq:a})--(\ref{eq:d}) are given in terms of the matrix elements of
$\hat{A}$.  We thus need to compute these matrix elements to determine
whether the perturbing operators in each limit are relevant or irrelevant.
As we have seen in Sec.~\ref{II}, this can be achieved for electron
systems with short-range interactions, i.e., the Hubbard chain 
at less than half-filling, and the
quantum wire in the low electron-density limit.  The dressed charges
$Z_{\nu \nu'}$ for the former case and the parameter $K_\sigma(m)$ for the
latter one can be calculated as functions of $m$ by solving the
corresponding Bethe ansatz integral equations.  We have solved the
integral equations numerically, and the results are discussed below.  The
readers who are interested in the details of the Bethe ansatz analysis
should refer to Refs.~\onlinecite{FrahmK1990} and \onlinecite{FrahmK1991}
for the Hubbard chain and Ref.~\onlinecite{KorepinBI1993} for the
Heisenberg chain.

\begin{figure}
\begin{center}
\noindent
\epsfxsize=0.45\textwidth
\epsfbox{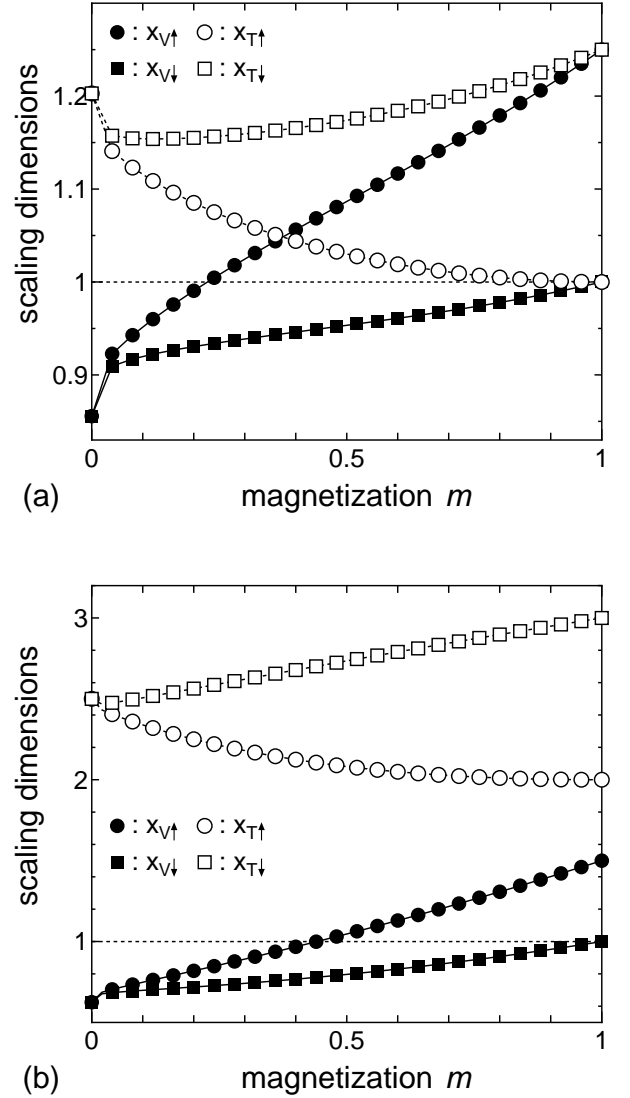}
\end{center}
\caption{
Magnetization dependence of scaling dimensions 
$x_{V \lambda}$ and $x_{T \lambda}$,
computed from the Bethe ansatz integral equations,
(a) for the Hubbard chain with $U/t = 4$ and $n = 0.5$ and
(b) for the quantum wire with $K_\rho = 1/4$ in the low-density limit.
}
\label{fig:sdim}
\end{figure}

\begin{figure*}
\begin{center}
\noindent
\epsfxsize=0.95\textwidth
\epsfbox{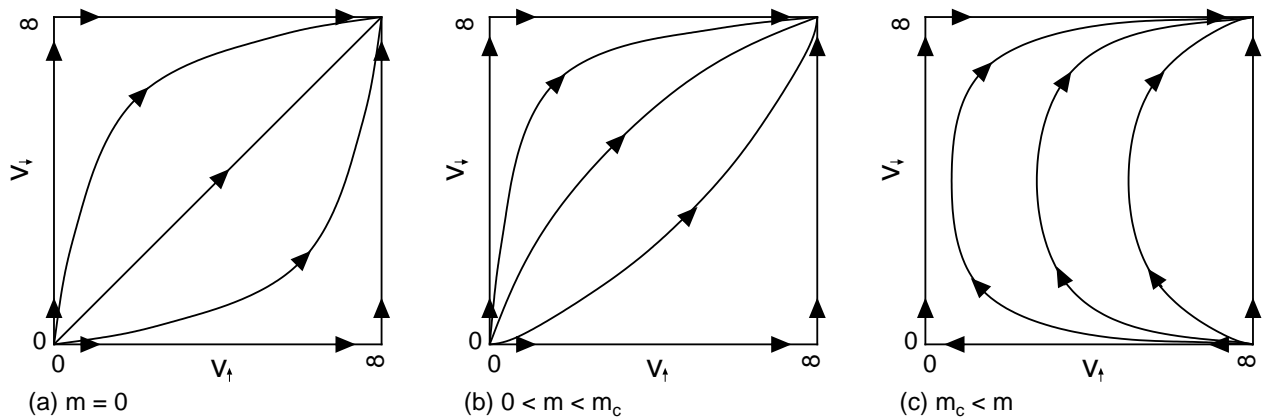}
\end{center}
\caption{
Schematic RG flow diagram for (a) $m = 0$, (b) $0 < m < m_c$, and 
(c) $m > m_c$.
}
\label{fig:RGflow}
\end{figure*}

In Fig.~\ref{fig:sdim} we plot the scaling dimensions
$x_{V \lambda}$ and $x_{T \lambda}$ obtained numerically
for the Hubbard chain at $n<1$ and the low-density quantum wire 
as functions of the relative magnetization $m$.
The two systems exhibit qualitatively the same behavior.
At zero magnetization $m = 0$,
we see that $x_{V \upa} = x_{V \dwa} < 1$ 
and $x_{T \upa} = x_{T \dwa} > 1$.
This is in accordance with the well-known scaling behavior 
in SU(2) symmetric systems, 
$\alpha_\upa^{\rm (a)}=\alpha_\dwa^{\rm (a)} < 0$ 
and $\beta_\upa^{\rm (b)} = \beta_\dwa^{\rm (b)} > 0$.
The dimensions $x_{V \upa}$ and $x_{V \dwa}$ increase with $m$.  In the
limit of full spin polarization $m\to1$ the dimension $x_{V \upa}$ reaches
a certain value $y$ greater than 1, whereas $x_{V \dwa}\to 1$.
(In the Hubbard model $x_{T\dwa}$ and $x_{T\upa}$ also approach
$y$ and 1 as $m\to1$.)
The most important point here is that the dimension $x_{V \upa}$ 
exceeds $1$ for $m$ larger than certain critical magnetization $m_c$.
This implies that the exponents $\alpha_\upa^{\rm (a)}$ 
and $\beta_\upa^{\rm (d)}$ change their sign at $m = m_c$, 
and the direction of the RG flows reverses.
The significance of this effect can be quantified by the value of 
$y$, which is given by
$y = 1+\left\{1-\frac{2}{\pi} \arctan[4 t\sin(\pi n)/U]\right\}^2$ 
for the Hubbard chain at $n < 1$ and 
$y = 1+2K_\rho$ for the quantum wire in the low electron-density limit.
On the other hand,
the dimensions $x_{T \upa}$ and $x_{T \dwa}$ are larger than 1
at any $m$, 
indicating that the scaling of the operators 
related to them does not change qualitatively
between $m>m_c$ and $m<m_c$.
We have checked that the dependence of the scaling dimensions on the
magnetization $m$ remains qualitatively the same regardless of the
interaction strength $U/t$ and the electron density $n < 1$ in the 
Hubbard model, or the exact value of $K_\rho < 1/2$ for low-density 
quantum wires.

From the magnetization dependence of the scaling dimensions $x_{V
  \lambda}$ and $x_{T \lambda}$ discussed above, we can deduce the RG flow
diagram of $(v_\upa, v_\dwa)$ as shown in Fig.\ \ref{fig:RGflow}.  When
the magnetization is small, $m < m_c$, the backscattering of electrons by
an impurity is enhanced by repulsive interactions.  
As a result, the RG trajectories go directly to the
strong-backscattering fixed point $(v_\upa, v_\dwa) = (\infty, \infty)$
[Fig.\ \ref{fig:RGflow}(a) and (b)].  The magnetic field bends the RG
trajectories upward but does not change the essential features of the
renormalization flow.

On the other hand, when the applied field is sufficiently strong to
achieve $m > m_c$, the backscattering operator $V_\upa$ of spin-$\upa$
electrons becomes irrelevant in the weak-potential limit.  That is, the
electron-electron interactions {\it suppress\/} the backscattering of
majority-spin electrons by a weak impurity.  As a result, the RG
trajectories in the vicinity of $(v_\upa, v_\dwa) = (0,0)$ flow toward the
line $v_\upa = 0$ [Fig.~\ref{fig:RGflow}(c)].  Thus, if the bare
backscattering amplitudes are not too large, they are renormalized toward
the asymmetric limit $(v_\upa, v_\dwa) = (0, \infty)$.  This means that,
in a certain regime of RG transformations or equivalently, at certain
energy and temperature range, a situation is realized where the impurity
potential becomes almost transparent for majority-spin electrons but
almost impenetrable for minority-spin electrons.  However, since the
asymmetric fixed point $(v_\upa, v_\dwa) = (0, \infty)$ is unstable, the
potentials are eventually renormalized to the strong-scattering fixed
point $(v_\upa, v_\dwa) = (\infty, \infty)$, with decreasing energy scale
or temperature $T \to 0$.  In this sense, a weak impurity potential can
have a spin-filtering effect generating a spin-polarized current.

\section{SUMMARY}
\label{IV}

In this paper we have studied the effect of magnetic field on the RG
flow of a single impurity potential in 1D interacting electron
systems.  Within the Abelian bosonization theory, low-energy physics
of the system is described as a Gaussian model with two independent
modes of bosonic excitations.  The coupling parameters $A_{ij}$ are
obtained from the Bethe ansatz for the 1D Hubbard model at less than
half-filling as well as for a quantum wire at the low electron density
limit.  Using these results, we have evaluated the scaling dimensions
of the impurity potential and tunneling operators, and determined the
RG flow of the potential amplitudes near the fixed points.  We have
found that the magnetic field can cause a drastic change in the RG
flow diagram.  While in a weak field the repulsive interactions always
enhance electron backscattering by impurities, in a sufficiently
strong field this effect is reduced, and the backscattering of
majority-spin electrons by a weak-impurity potential may even be
suppressed by the interactions.  This means that if the amplitude of
the bare potential is small, a spin-filtering phenomenon, in which
only the majority-spin electrons can transmit through the renormalized
potential, can be realized in a certain temperature regime in the RG
flow.

So far we have assumed that the band is less than half filled, $n<1$.
The results obtained at $n<1$ can be easily translated to the case
$n>1$ by performing the particle-hole transformation, $n_\lambda \to
1-n_\lambda$.  Suppose that the band is more than half-filled and
partially spin polarized, $0<m<1$.  Then, it is the majority-spin
holes, i.e., the spin-$\downarrow$ holes, whose backscattering is
suppressed in strong magnetic field.  In other words, transport of
charge carriers of majority-spin is enhanced by interactions.

The renormalizations of the impurity potential in the presence of magnetic
field have also been considered in a recent
preprint.\cite{kamide,Schmeltzer2002}
Although the authors of Ref.~\onlinecite{kamide} also found the regime in
which the weak backscattering of the majority-spin electrons is suppressed
by the interactions, their results differ significantly from ours.  Most
importantly, in Ref.~\onlinecite{kamide} this interesting regime occurs
either in the presence of spin-dependent interactions between electrons,
or when the interactions are attractive.  Both of these regimes are
unlikely to be realized in realistic experiments.  In contrast to our
work, the theory\cite{kamide} was based upon the treatment of 1D electron
systems in magnetic field developed in Ref.~\onlinecite{KimuraKAL1994}.
As we mentioned in Sec.~\ref{IIA}, the latter approach assumes weak
magnetic field, i.e., small relative magnetization $m\ll 1$.  In contrast,
our effect of suppression of weak impurities by interactions occurs at
sufficiently strong field, when $m\geq m_c\gtrsim 0.2$ (see
Fig.~\ref{fig:sdim}) and is expected in the realistic case of
spin-independent repulsive interactions.

\begin{acknowledgments}
  T.H. was supported by a Grant-in-Aid from the Ministry of Education,
  Culture, Sports, Science and Technology (MEXT) of Japan (Grant
  No.~16740213).  The work of A.F. was in part supported by NAREGI and a
  Grant-in-Aid for Scientific Research (Grant No.~16GS50219) from MEXT of
  Japan.  K.A.M. is grateful to RIKEN for kind hospitality.  The work of
  K.A.M. was supported by the U.S. DOE, Office of Science, under Contract
  No. \mbox{W-31-109-ENG-38}.
\end{acknowledgments}

\appendix

\section{Generalized eigenvalue problem of $\hat{\mathcal{H}} \hat{C}$}
\label{A}

In this Appendix we show how the matrix (\ref{eq:14}) can be brought
to the diagonal form (\ref{eq:diagH}).
Suppose that ${\bm \omega}_j$ and ${\bm \eta}_j$ are the right
and left eigenvectors of $\hat{\mathcal{H}} \hat{C}$, respectively, i.e.,
\bea \hat{\mathcal{H}} \hat{C} {\bm \omega}_j &=& u_j {\bm \omega}_j,
\label{eq:rightev} \\
({\bm \eta}_j)^T \hat{\mathcal{H}} \hat{C} &=& u_j ({\bm \eta}_j)^T.
\label{eq:leftev}
\eea
Then, the matrix $\hat{\mathcal{H}} \hat{C}$ is given by 
\be
\hat{\mathcal{H}} \hat{C} 
= \sum_j {\bm \omega}_j u_j ({\bm \eta}_j)^T,
\label{eq:matHC}
\ee
where ${\bm \omega}_j$ and ${\bm \eta}_j$ obey 
the biorthogonal condition,
\be
({\bm \eta}_j)^T {\bm \omega}_{j'} = \delta_{jj'}.
\label{eq:biorthog}
\ee
Since both $\hat{\mathcal{H}}$ and $\hat{C}$ are symmetric, i.e., 
$(\hat{\mathcal{H}})^T =\hat{\mathcal{H}}$ and $(\hat{C})^T =\hat{C}$,
Eq.\ (\ref{eq:leftev}) can be rewritten as 
\bea
(\hat{\mathcal{H}} \hat{C})^T {\bm \eta}_j 
= \hat{C} \hat{\mathcal{H}} {\bm \eta}_j = u_j {\bm \eta}_j.
\nonumber
\eea
Hence, ${\bm \eta}_j$ satisfies the relation 
\be
\hat{\mathcal{H}} \hat{C} \hat{C} {\bm \eta}_j 
= \hat{C}^2 \hat{\mathcal{H}} \hat{C}^2 {\bm \eta}_j 
= \hat{C} \hat{C} \hat{\mathcal{H}} {\bm \eta}_j 
= u_j \hat{C} {\bm \eta}_j,
\ee
where we used $\hat{C}^2 = \hat{1}$.
This means that $\hat{C} {\bm \eta}_j$ is a {\it right\/} eigenvector 
of $\hat{\mathcal{H}} \hat{C}$ and proportional to ${\bm \omega}_j$, 
\be
{\bm \omega}_j = c_j \hat{C} {\bm \eta}_j, 
\ee
where $c_j$ is a constant.
Furthermore, assuming that the Hamiltonian matrix $\hat{\mathcal{H}}$ is 
positive definite,\cite{PositiveDefinite}
it follows that $c_j$ and $u_j$ have the same sign,
\bea 
({\bm \omega}_j)^T \hat{C}\hat{\mathcal{H}}\hat{C} {\bm \omega}_j 
&=& u_j ({\bm \omega}_j)^T \hat{C} {\bm \omega}_j 
= u_j c_j ({\bm \eta}_j)^T {\bm \omega}_j 
\nn \\
&=& u_j c_j > 0.
\eea
We can therefore relate ${\bm \eta}_j$ to ${\bm \omega}_j$ as 
\be
{\bm \omega}_j = {\rm sgn}(u_j) \hat{C} {\bm \eta}_j
\label{eq:relationRL}
\ee
without any loss of generality.
Using Eqs.\ (\ref{eq:matHC}) and (\ref{eq:relationRL}), 
one finds that the right eigenvectors ${\bm \omega}_j$ 
satisfy Eq.\ (\ref{eq:diagH}), 
\bea
\hat{\mathcal{H}} &=& \hat{\mathcal{H}}\hat{C}\hat{C} 
= \sum_j {\bm \omega}_j u_j ({\bm \eta}_j)^T \hat{C}
\nn \\
&=& \sum_j {\bm \omega}_j u_j ({\bm \omega}_j)^T {\rm sgn}(u_j)
= \sum_j |u_j| {\bm \omega}_j ({\bm \omega}_j)^T.
\nn
\eea

The vectors ${\bm\omega}_j$ satisfy certain orthonormal conditions,
which can be derived as follows.
Suppose that $\hat{P}$ is a matrix of parity transformation 
exchanging the right- and left-moving fields.  Due to the 
parity symmetry of the Hamiltonian, the vectors ${\bm \omega}_j$ and 
$\hat{P}{\bm \omega}_j$ have the following property: 
if ${\bm \omega}_j$ is a right eigenvector of $\hat{\mathcal{H}}\hat{C}$ 
with an eigenvalue $u_j$, i.e.,
$\hat{\mathcal{H}} \hat{C} {\bm \omega}_j = u_j {\bm \omega}_j$, 
then $\hat{P}{\bm \omega}_j$ is another right eigenvector, with eigenvalue 
$-u_j$,
\be
\hat{\mathcal{H}}\hat{C} \hat{P} {\bm \omega}_j
= - \hat{P} \hat{\mathcal{H}}\hat{C} {\bm \omega}_j
= -u_j \hat{P} {\bm \omega}_j,
\ee
where we used the relations 
$\hat{P}\hat{\mathcal{H}}\hat{P}=\hat{\mathcal{H}}$
and $\hat{P}\hat{C}\hat{P}=-\hat{C}$.
Hence, we can classify the vectors ${\bm \omega}_j$ into two pairs, 
${\bm \omega}_{P c}$ and ${\bm \omega}_{P s}$, 
where the right and left movers in each pair are related as
${\bm \omega}_{L\nu} = \hat{P} {\bm \omega}_{R\nu}$.
Using these results and Eqs.\ (\ref{eq:biorthog}) and (\ref{eq:relationRL}), 
we find that the vectors ${\bm \omega}_{P\nu}$ obey the orthonormal
conditions (\ref{eq:P-orthogonal}).

\section{Matrix $\hat{A}$}
\label{B}

Here we outline the diagonalization procedure transforming the Hamiltonian
(\ref{eq:Hameff}) to the form (\ref{eq:tHam}) and express matrix $\hat A$
in terms of the velocities $u_{\uparrow,\downarrow}$ and the
coupling constants $\tilde{g}_{i\lambda,\perp}$.

We denote the elements of the vectors ${\bm \omega}_{P\nu}$ as
\be
{\bm \omega}_{R \nu}
=
\left(\begin{array}{c}
a_{\nu \upa}
\\
b_{\nu \upa}
\\
a_{\nu \dwa}
\\
b_{\nu \dwa}
\end{array}
\right)
,~~~
{\bm \omega}_{L \nu}
= \hat{P} {\bm \omega}_{R \nu}
=
\left(\begin{array}{c}
b_{\nu \upa}
\\
a_{\nu \upa}
\\
b_{\nu \dwa}
\\
a_{\nu \dwa}
\end{array}
\right).
\ee
Due to the parity symmetry of these vectors, 
the fields $\phi_\lambda$ and $\Pi_\lambda$ do not mix with each other 
under the transformation.
The relation between $\tilde{\phi}_\nu$ ($\widetilde{\Pi}_\nu$) 
and $\phi_\lambda$ ($\Pi_\lambda$) has the form,
\be
\left(\begin{array}{c}
\tilde{\phi}_c
\\
\tilde{\phi}_s
\end{array}
\right)
= \hat{A}_+
\left(\begin{array}{c}
\phi_\upa
\\
\phi_\dwa
\end{array}
\right)
,~~~
\left(\begin{array}{c}
\widetilde{\Pi}_c
\\
\widetilde{\Pi}_s
\end{array}
\right)
= \hat{A}_-
\left(\begin{array}{c}
\Pi_\upa
\\
\Pi_\dwa
\end{array}
\right),
\ee
where the matrix $\hat{A}_\pm$ is given by 
\be
\hat{A}_\pm =
\left(\begin{array}{cc}
A_{\pm c \upa} & A_{\pm c \dwa}
\\
A_{\pm s \upa} & A_{\pm s \dwa}
\end{array}
\right)
\ee
with $A_{\pm \nu \lambda} = a_{\nu \lambda} \pm b_{\nu \lambda}$.
Using the orthonormal conditions Eq.\ (\ref{eq:P-orthogonal}), 
one can find that the matrices $\hat{A}_\pm$ are related to each other as 
$(\hat{A}_+)^T = (\hat{A}_-)^{-1}$.
We thereby arrive at Eq.\ (\ref{eq:tphi-to-phi}) 
by identifying $\hat{A} = \hat{A}_-$.

From the eigenvalue problem Eq.\ (\ref{eq:rightev}), 
the linear equation system for $A_{\pm \nu \lambda}$ is obtained as
\begin{widetext}
\be
\left(\begin{array}{cccc}
0 & 0 & u_\upa + \tilde{g}_{4\upa} + \tilde{g}_{2 \upa}
& \tilde{g}_{4 \perp} + \tilde{g}_{2 \perp}
\\
0 & 0 & \tilde{g}_{4 \perp} + \tilde{g}_{2 \perp}
& u_\dwa + \tilde{g}_{4\dwa} + \tilde{g}_{2 \dwa} 
\\
u_\upa + \tilde{g}_{4\upa} - \tilde{g}_{2 \upa}
& \tilde{g}_{4 \perp} - \tilde{g}_{2 \perp} & 0 & 0
\\
\tilde{g}_{4 \perp} - \tilde{g}_{2 \perp} 
& u_\dwa + \tilde{g}_{4\dwa} - \tilde{g}_{2 \dwa} & 0 & 0
\end{array}
\right)
\left(\begin{array}{c}
A_{+ \nu \upa}
\\
A_{+ \nu \dwa}
\\
A_{- \nu \upa}
\\
A_{- \nu \dwa}
\end{array}
\right)
= u_\nu
\left(\begin{array}{c}
A_{+ \nu \upa}
\\
A_{+ \nu \dwa}
\\
A_{- \nu \upa}
\\
A_{- \nu \dwa}
\end{array}
\right).
\ee
\end{widetext}
By solving this eigenvalue problem, we obtain analytical expressions
for the renormalized velocities $u_\nu$,\cite{PositiveDefinite}
\bea
u_c &=& 
\sqrt{\frac{p_\uparrow + p_\downarrow 
    + \sqrt{(p_\uparrow - p_\downarrow)^2 + 4qr}}{2}},
\label{eq:ucele} \\
u_s &=& 
\sqrt{\frac{p_\uparrow + p_\downarrow 
    - \sqrt{(p_\uparrow - p_\downarrow)^2 + 4qr}}{2}},
\label{eq:usele}
\eea
and the elements of $\hat{A}_\pm$,
\bea
A_{+ \nu \upa} &=& \sqrt{
\frac{(u_\upa + \tilde{g}_{4\upa} + \tilde{g}_{2\upa})(u_\nu^2-p_\dwa)
      + (\tilde{g}_{4\perp} + \tilde{g}_{2\perp}) q}
     {u_\nu (2 u_\nu^2 - p_\upa - p_\dwa)}
},
\nn \\
\label{eq:Aplusup} \\
A_{+ \nu \dwa} &=& 
\frac{u_\nu^2 - p_\upa}{q}~ A_{+ \nu \upa},
\label{eq:Aplusdw} \\
A_{- \nu \upa} &=&
\left( 1 + \frac{u_\nu^2-p_\upa}{u_\nu^2-p_\dwa} \right)^{-1}
~ \frac{1}{A_{+ \nu \upa}},
\label{eq:A-up} \\
A_{- \nu \dwa} &=&
\frac{u_\nu^2 - p_\upa}{r}~
\left( 1 + \frac{u_\nu^2-p_\upa}{u_\nu^2-p_\dwa} \right)^{-1}~ 
\frac{1}{A_{+ \nu \upa}},
\label{eq:A-dw}
\eea
where 
\bea
p_\lambda &=& (u_\lambda + \tilde{g}_{4\lambda})^2 - \tilde{g}_{2\lambda}^2
             + \tilde{g}_{4\perp}^2 - \tilde{g}_{2\perp}^2,
\label{eq:elementp} \\
q &=& (u_\upa + \tilde{g}_{4\upa} + \tilde{g}_{2\upa})
      (\tilde{g}_{4\perp} - \tilde{g}_{2\perp})
\nn \\
&&+ (u_\dwa + \tilde{g}_{4\dwa} - \tilde{g}_{2\dwa})
      (\tilde{g}_{4\perp} + \tilde{g}_{2\perp}),
\label{eq:elementq}
\\
r &=& (u_\upa + \tilde{g}_{4\upa} - \tilde{g}_{2\upa})
      (\tilde{g}_{4\perp} + \tilde{g}_{2\perp})
\nn \\
&&+ (u_\dwa + \tilde{g}_{4\dwa} + \tilde{g}_{2\dwa})
      (\tilde{g}_{4\perp} - \tilde{g}_{2\perp}).
\label{eq:elementr}
\eea
Substituting Eqs.\ (6.12)-(6.17) of Ref.\ \onlinecite{PencS1993} 
into Eqs.\ (\ref{eq:ucele})-(\ref{eq:elementr}), 
we obtain the expression of matrix $\hat{A}$ 
for the Hubbard model [Eq.\ (\ref{eq:AforHub})].


\begin{thebibliography}{99}
\bibitem{PositiveDefinite}
The Hamiltonian matrix $\hat{\mathcal{H}}$ is 
positive definite and the velocities $u_\nu$ are real 
if
$(u_\upa + \tilde{g}_{4\upa} + \tilde{g}_{2 \upa})
(u_\dwa + \tilde{g}_{4\dwa} + \tilde{g}_{2 \dwa})
> (\tilde{g}_{4\perp} + \tilde{g}_{2 \perp})^2$
and 
$(u_\upa + \tilde{g}_{4\upa} - \tilde{g}_{2 \upa})
(u_\dwa + \tilde{g}_{4\dwa} - \tilde{g}_{2 \dwa})
> (\tilde{g}_{4\perp} - \tilde{g}_{2 \perp})^2$.
These are achieved if the parameters $\tilde{g}_{i \lambda,\perp}$ 
are not too large compared to the velocities $u_\lambda$.
We note that the conditions are satisfied 
in the Hubbard model for arbitrary values of the coupling 
$U/t$, electron density $n$, and magnetization $m$.
\end{thebibliography}
\end{document}